\documentclass[fleqn,12pt]{wlscirep}
\usepackage[utf8]{inputenc}
\usepackage[T1]{fontenc}
\usepackage{booktabs}

\usepackage{float}

\newcommand\MyBox[2]{
  \fbox{\lower0.75cm
    \vbox to 1.7cm{\vfil
      \hbox to 1.7cm{\hfil\parbox{1.4cm}{#1\\#2}\hfil}
      \vfil}%
  }%
}
\AtBeginDocument{%
  \providecommand\BibTeX{{%
    \normalfont B\kern-0.5em{\scshape i\kern-0.25em b}\kern-0.8em\TeX}}}

\newcommand\blfootnote[1]{%
  \begingroup
  \renewcommand\thefootnote{}\footnote{#1}%
  \addtocounter{footnote}{-1}%
  \endgroup
}

\title{From Asking to Answering:\\ Getting More Involved on Stack Overflow}

\author[a]{Timur Bachschi}
\author[b]{Anik\'o Hann\'ak}
\author[a]{Florian Lemmerich}
\author[c,d]{Johannes Wachs}

\affil[a]{Chair of Computational Social Sciences and Humanities, RWTH Aachen University}
\affil[b]{Department of Informatics, University of Zurich}
\affil[c]{Institute for Information Business, Vienna University of Economics and Business}
\affil[d]{Complexity Science Hub Vienna}

\begin{abstract}
Online knowledge platforms such as Stack Overflow and Wikipedia rely on a large and diverse contributor community. Despite efforts to facilitate onboarding of new users, relatively few users become core contributors, suggesting the existence of barriers or hurdles that hinder full involvement in the community. This paper investigates such issues on Stack Overflow, a widely popular question and answer community for computer programming. We document evidence of a ``leaky pipeline'', specifically that there are many active users on the platform who never post an answer. Using this as a starting point, we investigate potential factors that can be linked to the transition of new contributors from asking questions to posting answers. We find a user's individual features, such as their tenure, gender, and geographic location, as well as features of the subcommunity in which they are most active, such as its size and the prevalence of negative social feedback, have a significant relationship with their likelihood to post answers. By measuring and modeling these relationships our paper presents a first look at the challenges and obstacles to user promotion along the pipeline of contributions in online communities. 

\end{abstract}

\begin{document}
\maketitle
\section{Introduction}
Collaborative\blfootnote{Direct correspondence to: johannes.wachs@wu.ac.at} knowledge hubs like Stack Overflow and Wikipedia depend upon a diverse, large and active contributor community. Becoming a contributor in such platforms has been described as \emph{leaky pipeline}, i.e., a sequence of stages in knowledge consumption and production activities~\cite{shaw2018pipeline}. Users face barriers or hurdles to to from one stage to the next and indeed most users never graduate to the most involved forms of participation. For reasons of platform sustainability and effectiveness, it is important that many community members from diverse backgrounds and topical interests advance down the pipeline of participation.

\begin{figure}[!t]
\includegraphics[width=\textwidth]{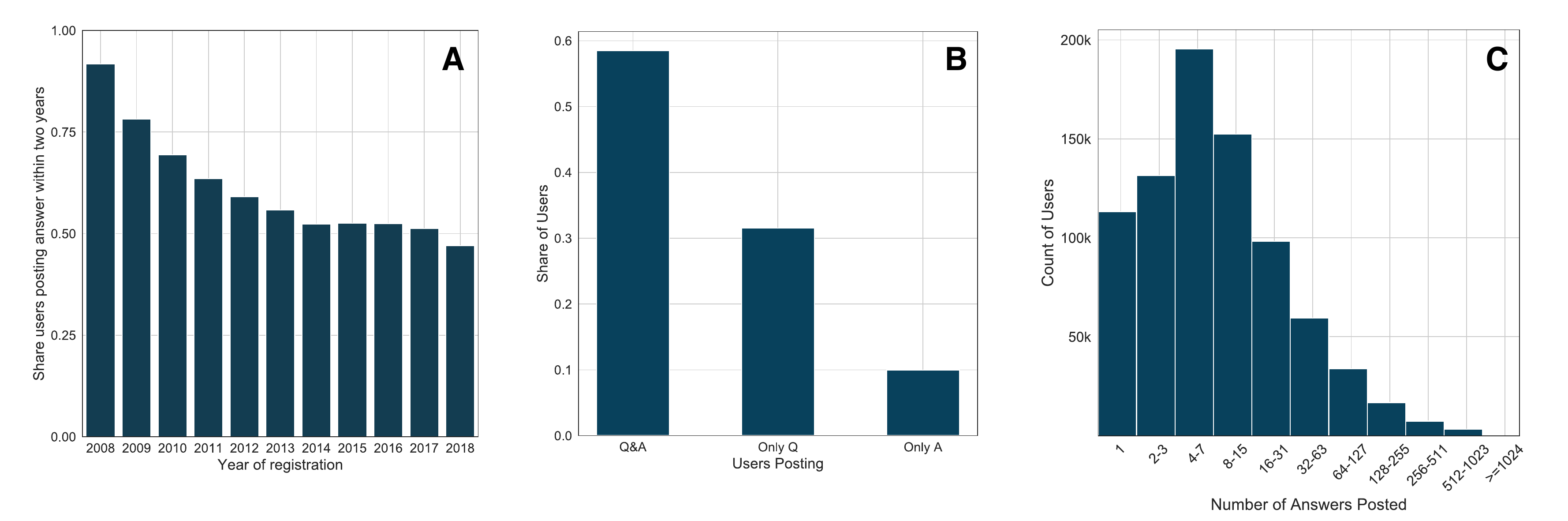}
\caption{Posting behavior of active users ($\geq5$ posts) on Stack Overflow. (A) The number of users posting an answer within two years of their registration is falling over time. (B) Nearly one-third of active users have only ever posted questions, but no answers. (C) Once a user posts an answer, they most of the time post multiple answers, suggesting that posting a first answer is a barrier or hurdle.}
\label{fig:overview}
\end{figure}

With that in mind, we study Stack Overflow, the web's largest Q\&A platform for software development and computer programming. There are growing concerns about the sustainability of Stack Overflow. There is a well documented gap between the number of content consumers and the number of contributors~\cite{mamykina2011design,srba2016stack}, and we also observe that many active users post only questions and not answers. In fact, only around half of all active users (which we define as having made least five posts) who joined the platform since 2014 post an answer within two years of registration (see Figure~\ref{fig:overview}A). Over time the share of users who have posted any answer at all has decreased: by now, users who never posted a single answer account for nearly one in three active users (see Figure~\ref{fig:overview}B). 

There is some reason to think that a user's first answer post is a difficult step to take. A significant majority of users who post one answer eventually post more answers (see Figure~\ref{fig:overview}C). We also find that users post more questions than answers early in their posting careers (see Figure~\ref{fig:transitionTimeDist}). Taken together, these descriptive findings suggest that there is an untapped pool of people with the potential to contribute answers. We need to understand better why this population seems to get stuck in this part of the pipeline of contribution. Even though researchers are aware of this sizeable community of potential contributors, little is known about the individual and community factors that predict whether a user will begin to post answers.

\noindent\textbf{Research Problem. }
In this paper we set out to understand if and when active users post an answer the first time. By doing so, we hope to identify potential barriers at an individual and at a (sub-)community level that are related to limited participation on Stack Overflow. Recognizing and dismantling such barriers supports platform sustainability by broadening and diversifying the contributor community, relieving the increasingly outnumbered group of core contributors and offering potential improvements to the knowledge base. 

\noindent\textbf{Approach and Results. }
We analyze a dataset of all posts on Stack Overflow with descriptive statistics and regression analysis. We find significant factors that can be associated with more frequent and quicker transitioning to posting answers to the platform. At a community level, we observe that subcommunities with higher negativity (high ratio of downvotes) correlates with later and fewer transitions to posting answers. People active in larger subcommunities are are less likely to post answers. On the individual level, we see that user tenure, gender, and geography are significantly related to answer posting outcomes. For example, users which we infer to be women are 52\% less likely to contribute any answers than similar men, and those who do post 12\% more questions before their first answer. Other significant factors at the individual level include tenure, geographic location, and tendency to post on weekends.

\noindent\textbf{Contribution and Implications. }
This paper describes to our knowledge the first study that analyzes a pipeline effect between contribution types, i.e., posting questions vs. answers, on the world's largest Q\&A platform. By carrying out a large scale data analysis, we identify factors that could constitute barriers for active users and inhibit more advanced involvement in collaborative knowledge platforms. The relationships between specific factors and the likelihood of progression through the pipeline suggest specific populations that may be encouraged to do more. Such efforts can improve the sustainability of platforms like Stack Overflow, and insure that different perspectives are present at all levels of contribution.

\section{Background and Related Work}

How best to organize the maintenance and expansion of public resources like collaborative knowledge hubs is a fundamental question of the social sciences~\cite{hardin1968tragedy}. The owners and designers of online knowledge databases in particular face many choices to manage their large and ever-changing communities. The institutional rules of a platform can shape long term outcomes in unexpected ways~\cite{ostrom2010beyond,keegan2017evolution}. Indeed, that even small site design choices are also known to make big differences in the patterns of participation~\cite{malik2016identifying} underscores the complexity of these environments. Two significant and interrelated issues that these communities face are sustainability and representativity. 

Online knowledge hubs face several challenges to their sustainability in the long run. As the growth of passive users and content outstrips the growth in core contributors with editing and moderation responsibilities, these contributors have more work to do~\cite{ortega2009wikipedia}. This growing pressure on what are essentially unpaid volunteers increases stress, hostility, and conflict~\cite{konieczny2018volunteer}, making it less appealing for new people to help. Not only do core contributors tend to contribute the majority of content on such platforms~\cite{mamykina2011design}, they also make important contributions to platform governance and organization~\cite{choi2018will}. Without restricting new content or users, platforms can improve this imbalance by improving core contributor retention and/or by onboarding new core contributors. 

It is also important that widely used knowledge hubs are built and maintained by people of many backgrounds. One reason is that culturally and intellectually diverse groups usually provide better solutions to tasks~\cite{page2008difference}, including, for example, writing high quality Wikipedia articles~\cite{arazy2011information}. Perhaps more important is the fact that gaps in knowledge bases arise when certain experiences are absent from the contributor base. For instance, the predominance of men on OpenStreetMap leads to an under-supply of information relevant to women~\cite{das2019gendered}. Such gender gaps, which have also been well-documented on Wikipedia~\cite{menking2015heart,wagner2015s} and Stack Overflow~\cite{vasilescu2013stackoverflow,ford2016paradise,may2019gender}, often intersect~\cite{fox2017imagining} with geographic~\cite{johnson2016not} and racial gaps. The significant differences in user behavior between the different language versions of Wikipedia only underscores the importance of building representative communities~\cite{lemmerich2019world}.

Stack Overflow in particular is an important platform to study from these perspectives. Besides the fact that we can learn about the general case from the specific (applying lessons to Wikipedia and online communities in general), Stack Overflow is an important part of the software community. Individuals use Stack Overflow to learn and developers seek solutions to problems they face at work~\cite{treude2011programmers}. There are even integrated development environment (IDE) plugins that allow developers to interface with Stack Overflow while coding~\cite{ponzanelli2013seahawk}. Stack Overflow provides a forum in which power users can answer questions about software libraries, relieving their owners and developers of a significant load~\cite{squire2015should}. It also plays a complementary role to API documentation~\cite{parnin2012crowd}. In the long run, it provides a platform for individuals to build their own confidence and expertise, creating a pipeline of future potential open-source software contributors~\cite{vasilescu2013stackoverflow}. In all, Stack Overflow is a key node in the social web of software development~\cite{storey2014r}.

While not all users come to platforms like Stack Overflow to become contributors~\cite{mamykina2011design}, several studies find that a significant number of users hesitate to participate because they feel they lack the necessary expertise~\cite{ford2016paradise,oliveira2018exchange}. It is likely that these feelings are especially salient when taking the step from asking to answering questions. A previous empirical analysis by Furtado et al.~\cite{mamykina2011design} supports this intuition: they find that users posting many questions in one time period are more likely to leave the site in the future than they are to post answers.

\section{Data and Features}
We use the Stack Overflow dump\footnote{https://archive.org/details/stackexchange}, accessed in July 2019, covering millions of posts made since 2008. Our primary focus is on the two most common kinds of posts that users make: questions and answers. As seen in Figure~\ref{fig:overview}, roughly one in three active users (i.e. users with at least five questions or answers) have never posted an answer. Among the roughly two-thirds of users who have posted an answer, some users post many questions before their first answers, while others ``jump right in'' by posting an answer as their first post. We report for summary statistics of the data in Table~\ref{tab:SODumpStats}. 

To better understand these differently participating populations of active users, we generate features to characterize them. The data contains information about individual users, including their profiles, posting histories, and the content of their contributions, including tags describing programming languages and frameworks in their posts. Using this data we create both user and community-level features that relate to our key outcomes. We focus on active users, which we define as those posting at least five questions or answers in total. While this excludes the majority of registered users (not to mention the even larger community of unregistered visitors to Stack Overflow), we argue that previous work has focused on why users take the first steps from registering to making initial contributions. Our contribution will focus on deeper involvement.

\begin{table}[H]
	\centering
	\begin{tabular}[ht!]{| l | r |}
		\hline			
		\#Posts               & 44,945,355 \\
		\#Answers             & 27,107,580 \\
		\#Questions           & 17,738,809 \\
		\#Other posts         & 99,066     \\
		\#Users with 5+ posts (Active Users) & 1,188,419  \\
		\#Posts by Active Users & 37,617,578  \\
		\hline  
	\end{tabular}
	\caption{Statistics for the Stack Overflow Dataset.}
	\label{tab:SODumpStats}
\end{table}

\subsection{Dependent Variables}
Our primary outcomes of interest are whether or not an active user posts an answer to a question on Stack Overflow, and how quickly such a post occurs in their posting history. In the latter case, we count the number of questions a user posts before their first answer among users who ever post an answer. The correlation between the count of questions before the first answer and the time between the two events are highly correlated (Spearman's $\rho = 0.91$) The binary outcome of posting an answer and the count of questions until a user's first answer, are our key dependent variables. 

\begin{figure}[!ht]
\includegraphics[width=\textwidth]{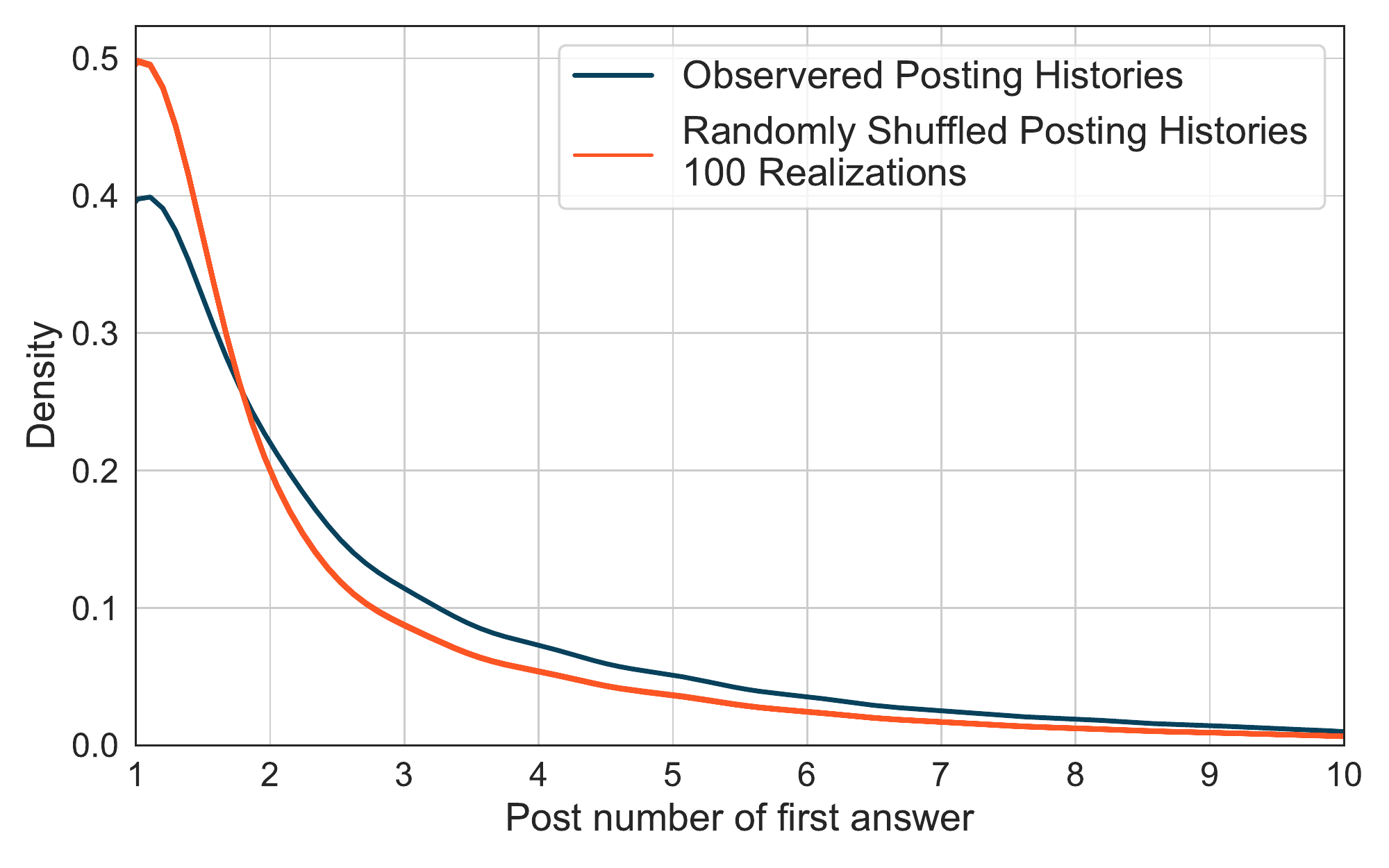}
\caption{We plot the distribution of which post is a user's first answer, for active users on Stack Overflow in dark blue. For about 40\% of users the first post is an answer. The average active user (by mean) posts an answer after 3.8 posts (median: 2). In red we plot 100 realizations of the same distribution under a shuffled null model. The null model randomizes the order of question and answer posts made by each user. The null model's curve is significantly steeper, suggesting that users tend to post more questions before their first answer than expected if their posting behavior were random.}
\label{fig:transitionTimeDist}
\end{figure}

Roughly 375,000 active users (31.5\%) have not posted an answer. Among those that do post an answer, a majority of users post one within their first few posts, though a significant share of users only post their first answers after several questions. We plot the empirical distribution of the rank of users' first answer posts in Figure~\ref{fig:transitionTimeDist}. In the same figure we also plot 100 realizations of a counterfactual distribution, generated by randomizing the order of posts in the posting career of each user. This null model simulates what we would expect the distribution of the rank of users' first answers to look like if users had posted their questions and answers in random order.

The plot indicates that users tend to post their first answer later in their posting histories than we would expect if the order of their posts were random. This tendency to ask before answering, complementing our earlier observations that many users either never post answers or post multiple answers, suggests that there is something special about a user's first answer. Whether this step is conceptualized as a hurdle or barrier or alternative choice, we would like to quantify if user or subcommunity level features are related to the likelihood that a user takes this step, and if so, how quickly they will do so.

\subsection{User features}
We begin by focusing on individual aspects of users that may predict if and when they post answers. These features include the level of detail a user provides about themselves, the length of their tenure, their location, their gender, and whether they post more frequently on weekends.

\subsubsection{Profile}
Users on Stack Overflow have personal pages called profiles where they can reveal personal information about themselves and link to their identities on other platforms such as Twitter or GitHub. Such profiles are ways for users to share more about themselves, building trust with the broader community~\cite{erickson2000social} and linking their digital selves to their real identities. Users providing such information on platforms may be seen as more reliable or reputable - for example articles on Wikipedia written by users disclosing significant personal information are much more likely to become ``featured'' articles~\cite{stein2007does}. Evidence from studies on GitHub suggests that users actively manage their public image on the website~\cite{dabbish2012social}. It seems that users are also leveraging their user profiles and accomplishments on Stack Overflow to find work: a recent study finds that activity on Stack Overflow falls significantly after a user starts a new job \cite{xu2020makes}, though this may be because they have less time to devote to the platform.

We focus on whether or not users fill out their ``About Me'' section. This is a free-form text field in which users are encouraged to write a few words about themselves. Unlike other fields in the profile such as links to other social media platforms, this field is particularly oriented towards allowing the user to introduce themselves to the Stack Overflow community. We expect that users who take the time to fill out their profiles are more likely to have transitioned to posting answers, and that they will have done so more quickly. We note here that we only have information from this field at the time of the most recent data we use, from 2019. As a result, it may be the case that users who post answers subsequently fill out their profiles, seeking to tie their personal identities to their Stack Overflow profiles because of their investment in the platform.

\subsubsection{Year of registration}
Stack Overflow has been online since 2008. Following a brief private beta, the site was opened to the public. Since then more users register each year. Many online communities have conflicts between long-time users and newcomers~\cite{danescu2013computational}. Newcomers to Stack Overflow have suggested that it is becoming harder to answer questions as the low-hanging fruit have been picked and there is a large population of users who quickly answer questions~\cite{bosu2013building}. More established users lament what they perceive to be a decline in posting quality of new users: for example, so-called \textit{help vampires} who ask low quality questions which could easily be resolved with a little effort from the poster \cite{srba2016stack}. Indeed a secular rising trend of unanswered, deleted, and moderated questions suggests that the average quality of posts on Stack Overflow is falling. To understand the relationship between a user's tenure on the site and if and when they post an answer, we record each user's year of registration. We hypothesize that more recently registered active users are less likely to post answers, and if they do are likely to ask more questions before their first answer.

\subsubsection{Geography}
While users of Stack Overflow come from all over the world, the distribution of users is highly skewed towards North America and Western Europe~\cite{oliveira2016participation}. Previous work highlights several factors that may explain this outcome. One factor is that Stack Overflow has an English-language policy \cite{non-english}, creating barriers for non-native English speakers using the platform \cite{guo2018non,SO_statistics}. There are also localized versions of Stack Overflow in Russian and Portuguese, which may divert users from the more widely-used primary platform. Beyond linguistic barriers, previous work suggests differences in cultural attitudes leads to differences in behavior on Q\&A platforms \cite{yang2011culture}. It is also likely that unequal access to digital infrastructure across and within countries influences who accesses Stack Overflow and how they participate on the platform \cite{chen2004global}. 

All of these factors likely play a role in user advancement on Stack Overflow. We therefore infer user location from free-text in their profiles using the Python library \textit{Geotext}\footnote{https://github.com/elyase/geotext}, an open source software that detects mentions of locations in text. This approach has several limitations, for example that the library is biased towards English-language representations of place names (i.e. ``Germany'' instead of ``Deutschland''). While we were only able to infer location for 47\% of users, the distribution of users across countries was similar to data reported in the 2019 Stack Overflow survey. The US (inferred: 22\%, survey: 23.6\%), India (inferred: 17.8\%, survey 10.2\%), the UK (inferred: 6\%, survey: 6.5\%), Germany (inferred: 4.9\%, survey: 6.6\%), and Canada (inferred: 3.5\%,survey: 3.8\%) were the top five countries in both rankings. As most previous literature suggests that the greatest geographic disparities in participation online and on Stack Overflow specifically occur between the developed and developing world, we simplify by categorizing each user as coming from a country in either the so-called global north or global south as classified by the Wikimedia Foundation\footnote{https://meta.wikimedia.org/wiki/List\_of\_countries\_by\_regional\_classification}.

\subsubsection{Gender}

Differences in online participation between men and women are widely studied~\cite{fatehkia2018using,garcia2018analyzing}. Much work has been done to measure gender gaps and barriers in participation and achievement in computer programming and software~\cite{dias2019barriers}, including Stack Overflow ~\cite{vasilescu2013stackoverflow,may2019gender}. Women tend to post more questions than answers, and tend to mention different kinds of obstacles to making contributions~\cite{ford2016paradise}. One hypothesized force behind the gap is a disparity in confidence between men and women~\cite{peterson2010gendered}. A recent study suggests that these effects go beyond individuals: women are more likely to engage in a post if they observe other women active on the thread~\cite{ford2017someone}. 

As previous research indicates that women tend to pose more questions and fewer answers, we expect that women are less likely to post answers at all, and to post more questions before their first answer. Deviations from these expectations, for example that among users who do post answers women are not any slower to do so than men, would signal that there are specific points in the pipeline that are more significant hurdles than others.

As Stack Overflow does not ask users about their gender, we must infer gender from user profiles. Following previous work~\cite{vasilescu2013stackoverflow} we infer user gender from usernames and location using a dictionary-based approach. In particular, we apply the \textit{Gender-Guesser} software\footnote{https://github.com/lead-ratings/gender-guesser}, which uses regional and national dictionaries recording the frequency that men and women have a given surname. We attempt to infer gender only for those users for whom we can infer location to improve the accuracy of the gender inference. We apply the tool both on the full user name and on its first token, and only consider those users for which the tool claims high confidence. In this was we classify roughly 36.5\% of users with a location as likely men or women. Among these users, 7.2\% are classified as likely women - a number in line with previous work and the Stack Overflow survey (7.6\%) ~\cite{SO_statistics}.

This approach makes several significant simplifying assumptions: that gender is a binary phenomenon and that it can accurately be inferred from names. We also assume that the error rate between the classes is balanced. We discuss how these limitations may lead to bias in our results in the conclusion of the paper.

\subsubsection{Weekenders}
Even though a significant share of Stack Overflow users code as a hobby~\cite{SO_statistics}, 87\% of traffic to the site occurs on weekdays~\cite{silge2018weekend}. Users posting on weekends may be more likely to be hobby programmers, learners, or pursuing side projects: this is reflected in significant differences in the prevalence of certain programming languages between weekends and weekdays~\cite{allamanis2013and}. Open-source software developers are likely overrepresented among weekend posters - roughly a third of them work primarily on nights and weekends ~\cite{claes2018programmers}. Not only do such developers likely have different motivations and interests, they are likely to engage with projects or communities for different reasons~\cite{miller2019people}. These differences manifest, for example, in the finding that questions posted on weekends are more likely to be answered~\cite{bosu2013building}. As these unobserved differences in motivation and engagement are likely to influence whether a user will post an answer, we include a feature capturing whether a user frequently posts on the weekend. Specifically, a user is classified as a frequent weekend poster if at least 2/7ths of their posts are on weekends.

\subsection{Community features}
Individual users have many distinguishing traits but are also embedded in distinct communities around programming languages and frameworks with their own norms and attitudes. Indeed heterogeneity in activity amongst these sub-communities has been used to forecast programming language and framework popularity~\cite{chen2016mining} or to identify user areas of expertise~\cite{menshikova2018evaluation}.

Different kinds of users can be categorized into communities based on unobserved attributes such as personal taste for programming languages or the technologies they use at work or school. Different aspects of these communities likely provide a more or less hospitable environment for users willing and able to begin posting answers. These communities have significantly different structural features that may influence the behavior of its members, for instance the age of the community, its recent growth, its size in terms of active users, and emergent cultural factors such as the prevalence of negative social feedback within the community.

We assign users to communities of specialization on Stack Overflow by using tags. Tags are categorizations used to label and filter questions on the platform. For example the ``python'' tag, one of the most popular tags, indicates that a question has to do with Python. Answers inherit the tags of the questions they address. We note that questions can have multiple tags and that there are many very rarely used tags. To reduce noise, we track only the top 500 most frequently used tags, grouping the rest into an ``other''. We sort users into tag communities by considering the most frequent tag used on questions up to and including their first answer post. Less than 1,000 out of over 1 million users are categorized in the ``other'' tag community. The user inherits the tag-community features we describe in the following subsections. In the case that a user has two or more equally used most frequent tags, we average the community-level features across these specializations.

\subsubsection{Community Size}
We count the number of users in our sample that have specialization in that community. Larger communities may be more competitive, with many users watching the queue of questions, making it more challenging to be the first to answer a question. Users in larger communities may encounter the same posters more rarely than in smaller communities, decreasing the chance that users will build ties and feel that they are a part of an actual community of users. There is also a possibility that programming novices begin their careers in larger communities, as more popular languages and frameworks tend to have more resources. ``Javascript'', ``java'', ``php'' and ``python'' are some of the largest communities, while ``dart'', ``clojure'' and ``neo4j'' are among the smallest we consider in our analysis.

\subsubsection{Community Negativity}
Many users have expressed concerns about negativity or hostility of other users on Stack Overflow~\cite{so_welcoming}. We measure \textit{community negativity} as the ratio downvotes, user actions expressing disapproval on posts, to the total number of votes cast on posts in the community:

\begin{equation}
CN = \frac{\#downvotes\;in\;subcommunity}{\#votes\;in\;subcommunity}
\end{equation}

Tag communities with exceptionally low levels of negativity include ``emacs'', ``git'', ``ruby-on-rails'', and ``clojure''. Communities with high levels of negativity include ``excel'', ``vba'', ``php'', and ``arduino''. The high negativity of the ``excel'' and ``vba'' communities may be because of an overrepresentation of users asking and answering questions about software they use at work~\cite{silge2018weekend}.

\subsubsection{Share of Reputation Awarded in last year}
Different programming languages and frameworks are falling in and out of popular favor every year. New users in a relatively popular community may face a more hectic pace and anonymous environment than those joining a more established and steady one or even a community on the decline. We distinguish between such communities by calculating the share of reputation points awarded to posts carrying that community's tag made in the past year. Reputation points are Stack Overflow's way of recognizing valuable contributions. 

Communities with a high share of reputation awarded recently include ``keras'', ``kotlin'', ``vue'', and ``dart'' - all examples of very young programming languages or frameworks. Communities with a low share of reputation awarded recently include ``silverlight'', ``backbone-js'', ``flash'', and ``svn''.

\subsubsection{Concentration of Reputation}
Communities with a distinguished elite may be less accessible to new individuals who want to become contributors. On the other hand, such a group may present ideal mentors or role models for new users. For each community we calculate the distribution of reputation points awarded to users, measuring its concentration by the share of reputation won by the top 10\% of users in each community. The tags ``arduino'' and ``unity-3d'' have a low concentration of reputation among the top 10\% of their users, while mainstream languages including ``python'', ``c++'', and ``java'' have highly concentrated reputation scores.

\subsubsection{Features of Large Communities}
To facilitate interpretation of our community features, we report the top and bottom five communities according to each feature, considering only programming-language communities from the top 25 tags, in Table~\ref{tab:bigcommunitystats}. Even among these large communities, there is significant variation. There are nearly four times as many down votes cast on ``php'' posts than there are on ``.net'' posts, proportionally speaking. The total share of reputation awarded on posts with the tag ``python'' in the last year is four times the amount awarded on ``.net'' tagged posts. The concentration of reputation within the top 10\% is nearly 90\% for ``c++'', and only 77\% for ``ajax''. These differences suggest that posters in different subcommunities on Stack Overflow have different experiences. We will soon see that these structural differences are significantly related to different outcomes in user posting behavior.
\begin{table}
\centering
\begin{tabular}{|lp{20mm} || lp{25mm} ||lp{30mm} |}
\toprule
     Tag &    Community Negativity & Tag & \% of Rep. of Community in Prev. Year &Tag & Top Decile User Rep. Share in Community.\\
\midrule
     php &  0.19 &    python &            0.08 &      c++ &              0.89  \\ 
       c &  0.18 & node.js &            0.07 &python &              0.89 \\
   mysql &  0.16 & javascript &            0.05&  c\# &              0.88  \\
    html &  0.15 & c++ &            0.05 & java &              0.88 \\
     sql &  0.13 &      sql &            0.05& javascript &              0.87  \\
     			\hline 
 android &  0.10 & database &            0.03&   ios &              0.83\\
 asp.net &  0.08 & ajax &            0.03& node.js &              0.83 \\
     ios &  0.08 & jquery &            0.02& asp.net &              0.82  \\
 node.js &  0.06 & asp.net &            0.02& database &              0.81 \\
    .net &  0.05 &  .net &            0.02& ajax &              0.77 \\
\bottomrule
\end{tabular}
	\caption{The top and bottom five tags from the top 25 most frequently used programming-language or framework related tags, ranked by our community measures, i.e., community negativity, the share of reputation points within a community awarded within the last year, and the concentration of reputation in the top 10\% of users.}
	\label{tab:bigcommunitystats}
\end{table}

\section{Analysis}
We now proceed with our analysis. First, we investigate correlations between user and community features and our key dependent variables in a descriptive analysis. We then apply a multiple regression modeling framework to quantify the relationship between our features and outcomes while controlling for several possible confounding factors. We interpret both the statistical significance and estimated effect sizes of our features.

\subsection{Correlations}
There are several interesting correlations between the user and community features and our outcomes of interest which we report in Table~\ref{tab:correlations}. For example, users that disclose more information about themselves are more likely to contribute answers, and are more likely to do so quickly. Users in growing communities and communities with a strong elite are significantly less likely to contribute answers, and when they do so post more questions before their first answer. These correlations are merely suggestive, as confounding factors may be at play.
\begin{table}[H]
	\centering
	\begin{tabular}[ht!]{| l | r | r|}
	\hline
	    Feature  & User Posts Answer & Questions until First Answer \\
		\hline			
		Account age               & .23& .01 \\
		Shares Personal URL           &.25  & -.12 \\
		Shares AboutMe          & .28&  -.09\\
		Inferred Woman        & -.11   & .04 \\
		Frequent Weekend Poster        & -.06  & .01 \\
		From Global North        & .04  & -.04 \\
		Community Negativity       & -.09   & .05 \\
		Users in Community (log) & -.02& .03 \\
		\% of Rep. of Community in Prev. Year & -.17& .12  \\
		Top Decile User Rep. Share in Community &  -.27& .34  \\
		\hline  
	\end{tabular}
	\caption{Spearman correlations between the user and community features and the dependent variables. All correlations are significant at p
	$<$ .0001.}
	\label{tab:correlations}
\end{table}

\subsection{Models}
In order to understand how the relationships between our features and dependent variables mediate each other, we employ multiple regression models. For the binary outcome of whether a user posted an answer we employ logistic regression, and for the count outcome of how many questions a user posts before their first answer we employ negative binomial regressions. We run four regressions in all, two for each dependent variable. In both cases we fit a model on the smaller dataset of users for which we could infer location and gender, and a model on the whole population of users in which we drop the geographic and gender variables.

We also include fixed effects for a user's most commonly posted tag, capturing their specialty. These controls capture any additional variation across user communities that the features we generate may have missed.

\subsubsection{Results}
In this section we present and interpret the results of our regression models. We find that several individual and sub-community level features are significant predictors of both whether a user posts answers and how quickly they do so. We present our main findings in Table~\ref{tab:modelresults}. We measure model fit with McFadden's Pseudo $R^{2}$, finding that our models explain roughly a third of variance in a user's number of questions posted before a first answer and one fifth of the variance in the likelihood they post an answer at all.

\begin{table}[!htbp] \centering
	\setlength\tabcolsep{1.5pt}
	\resizebox{\textwidth}{!}{
		\begin{tabular}{@{\extracolsep{1cm}}lp{1.7cm}p{1.7cm}p{1.7cm}p{1.7cm}}
			\setlength\tabcolsep{1.5pt}
			\\[-1.8ex]\hline
			\hline \\[-1.8ex]
			& \multicolumn{4}{c}{\textit{Dependent Variables}} \\
			\cline{2-5}
			\\[-1.8ex] & \multicolumn{2}{l}{\# of Questions before First Answer (NB)} & \multicolumn{2}{l}{Posts Answer (Logit)} \\
			                       &                  &                  &                  &                  \\[-1.8ex] & (1) & (2) & (1) & (2) \\
			\hline \\[-1.8ex]
			Intercept              & $-$8.029$^{***}$    & $-$7.503$^{***}$    & 2.493$^{***}$    & 1.629    \\
			                       & (0.064)          & (0.148)          & (0.126)          & (0.403)          \\
			                       &                  &                  &                  &                  \\
			Account Age (Years)  & 0.048$^{***}$ & 0.050$^{***}$ & 0.165$^{***}$ & 0.258$^{***}$ \\
			                       & (0.003)          & (0.001)          & (0.001)          & (0.003)          \\
			                       &                  &                  &                  &                  \\
			Website URL on profile & $-$0.200$^{***}$ & $-$0.133$^{***}$ & 0.957$^{***}$    & 0.542$^{***}$    \\
			                       & (0.003)          & (0.007)          & (0.008)          & (0.018)          \\
			                       &                  &                  &                  &                  \\
			Filled-out AboutMe     & $-$0.080$^{***}$ & $-$0.062$^{***}$ & 1.305$^{***}$    & 0.953$^{***}$    \\
			                       & (0.003)          & (0.006)          & (0.006)          & (0.015)          \\
			                       &                  &                  &                  &                  \\
			Inferred Woman         &                  & 0.115$^{***}$    &                  & $-$0.727$^{***}$ \\
			                       &                  & (0.012)          &                  & (0.023)          \\
			                       &                  &                  &                  &                  \\
			 Frequent Weekend Poster           & $-$0.025$^{***}$ & 0.004    & $-$0.198$^{***}$ & $-$0.174$^{***}$ \\
			                       & (0.003)          & (0.008)          & (0.005)          & (0.017)          \\
			                       &                  &                  &                  &                  \\
			From Global North      &                  & $-$0.047$^{***}$ &                  & $-$0.091$^{***}$ \\
			                       &                  & (0.007)          &                  & (0.016)          \\
			                       &                  &                  &                  &                  \\
			Community Negativity   & $-$0.935$^{***}$    & $-$0.717$^{***}$    & $-$0.370$^{**}$ & $-$0.146 \\
			                       & (0.090)          & (0.208)          & (0.171)          & (0.543)          \\
			                       &                  &                  &                  &                  \\
			Users in Community (log)   & 0.222$^{***}$    & 0.227$^{***}$    & $-$0.269$^{***}$ & $-$0.148$^{***}$ \\
			                       & (0.001)          & (0.010)          & (0.002)          & (0.007)          \\
			                       &                  &                  &                  &                  \\
			\% of Rep. of Community in Prev. Year   & 1.019$^{***}$    & 0.979$^{***}$    & 0.012 & 0.038 \\
			                       & (0.092)          & (0.208)          & (0.157)          & (0.442)          \\
			                       &                  &                  &                  &                  \\
			Top Decile User Rep. Share in Community   & 7.335$^{***}$    & 6.612$^{***}$    & 0.791$^{***}$ & 0.797$^{*}$ \\
			                       & (0.072)          & (0.166)          & (0.139)          & (0.441)          \\
			                       &                  &                  &                  &                  \\
			\hline \\[-1.8ex]
			Observations           & 813,365          & 155,263          & 1,188,415        & 183,812          \\
			McFadden Pseudo R$^{2}$         & 0.353            & 0.363            & 0.205            & 0.213            \\
			\hline
			\hline \\[-1.8ex]
			Significance thresholds:  & \multicolumn{4}{l}{$^{*}$p$<$0.1; $^{**}$p$<$0.05; $^{***}$p$<$0.01}. \\
			Note: User community fixed-effects included.
		\end{tabular}
	}
	\caption{Models predicting the number of questions posted before a user's first answer post (negative binomial regression) and the likelihood that a user ever posts an answer (logistic regression) to posting answers. Each model is fit first to the full dataset, and then to the sub-population of users for which we can infer gender and location.}
	\label{tab:modelresults}
\end{table} 

The results demonstrate that individual social features and community attributes have strong relationships with the likelihood that a user posts answers. For instance, an additional year of tenure on the platform is related to $e^{0.048}-1 \approx 0.049 = 4.9\%$  more questions posted before a first answer, and a 17.9\% higher likelihood of posting an answer at all. The latter finding supports the motivating impression from earlier in the paper that newer users are less likely to eventually post answers. Yet among those new users that do post answers, they tend to post their first answer
after fewer questions. This underlines the importance of considering both models: considering when answering users post their first answer ignores the significant pool of users that never post an answer at all.

Participating in the social life of Stack Overflow is also an important predictor of posting answers. Users who link to their personal pages are more likely to post answers (160\%) and do so after 13\% fewer posts. Similarly users who fill out their about me post answers 7.7\% sooner and are over 260\% more likely to do so at all.  The relationships between weekend posting and the dependent variables are much less extreme: weekend posters are 18\% less likely to post answers, yet post answers 2.5\% sooner on average.

Turning to the community features, we find several important relationships between attributes of the communities users are embedded in and their behavior. A 10\% increase in community negativity, for example, decreases the number of questions posted before a user's first answer by roughly 9\%, and decreases the chance the user posts answers at all by 3.6\%. Users in larger communities tend to post their first answer later, and are less likely to do so at all. The recent popularity of a community increases the time to first answer but has no significant relationship with whether or not a user posts an answer. Finally, an increase in the concentration of reputation among the top decile of users in a community is strongly related to an increase how long it takes to post an answer, but also the likelihood of answering overall.

Most of the significant relationships described above hold in the models fit to data on users for which we could infer gender and location. These two features also present interesting results: inferred women users are roughly 52\% less likely to post answers, and do so 12\% more slowly than their inferred man counterparts. We interpret these findings as suggesting that the gendered pipeline on Stack Overflow has significant leaks at several stages. Users from the global north are less likely to post answers, but when they, do so more quickly.

\section{Discussion}

In this work we analyzed individual and community level factors predicting whether and how quickly an active user contributes to Stack Overflow by answering questions. We review our findings and discuss their implications, then consider limitations and potential future work.

Our primary models explored the relationship between individual attributes of users and the subcommunities they occupy, and the likelihood that they will post a first answer. As we discussed in the introduction, there is a trend that new users are less likely to ever post answer and there is a growing group of active users who have posted only questions. There is also evidence of a barrier effect: most users who post an answer post more than one. Our analysis therefore zooms in on the transition active users make to posting answers. We model both the likelihood of a user posting an answer, and, among the population of answer posters, how quickly in their posting career the first answer comes. Our regression model framework allows us to consider the relationships of multiple features of users simultaneously, holding variation of the other features constant.

We find numerous significant relationships between individual and community factors, and our dependent variables. Users with older accounts, filled out profiles, and those coming from communities with a high concentration of reputation score are more likely to post answers. Women, weekend posters, users from the global north, and users from large and more negative communities are less likely to post answers. In most cases the features that predict a greater likelihood of posting an answer, are also correlated with answering earlier. For instance, women tend to post more questions before they post their first answer. There are exceptions, however: users from the global north, who are less likely to post an answer, ask fewer questions before their first answer. Similarly, even though users from communities with unequal reputation score distributions are more likely to post answers, they take significantly longer to do so.

The magnitudes of these relationships are sometimes surprisingly high. All else equal, women are 52\% less likely to post an answer. This echoes previous work on gender differences in posting behavior on Stack Overflow, which finds that men tend to post more answers than women, and that barriers may have different effects on different groups of users~\cite{ford2016paradise}. That answer-posting women tend to ask more questions before their first answer than men provides a bit more information on the potential roots of the observed behavioral differences. Perhaps question asking serves as a way to build confidence. 

The relationship between answer posting and geographic location is also interesting. A recent analysis of the Stack Overflow user survey~\cite{SO_statistics}, found that a significant number of users referenced the English language when discussing reasons why they hesitate to contribute to the platform. We have already mentioned that Stack Overflow has non-English language versions~\cite{guo2018non}. These platforms are a fraction of the size of the main Stack Overflow site, limiting their potential to benefit from network effects. In the future, Stack Overflow may consider employing machine translation to facilitate cross-language interactions. This would improve accessibility and bring more contributors to the platform - though poses significant technical challenges.

The significant relationship between community level features and answering activity is evidence that user behavior depends on their interactions and experiences with others on the platform. It is unclear from our analysis to what extent users self-select into communities matching their style, and how much of the differences are due to emergent culture of the communities themselves. Likely both factors are at play: while some communities may have well-defined images that attract like-minded newcomers, many programmers, novice or expert, use specific programming languages for reasons beyond their control. For instance a computer science student at a certain university may learn Java because that is what is required in the introductory programming course. A young woman seeking to learn programming in a friendly environment might attend a Django Girls or Rails Girls event, and become a lifelong Python or Ruby programmer, respectively. 


Our findings have several practical implications. First and foremost, we emphasize that posting a first answer is a significant barrier for many users. Hence many of the interventions designed to help new users post their first questions, from the mentors available in special chatrooms~\cite{ford2018we} to the tips and hints that appear on screen when a user types their first question ~\cite{calefato2018ask}, could be extended to help users post their first answer. At the same time, we recognize that posting an answer is a fundamentally different process than asking a question. To answer a question, a user must find a question they can answer, and answer it in a timely manner, as duplicate answers are generally frowned upon. No such search or notion of speed plays a direct role in the process of asking a question. Therefore, getting people to post their first answer effectively likely requires different kinds of support and encouragement. One scenario in which posting an answer has less time pressure is when a question is already answered. New answer posters may be encouraged to revisit answered posts if they have an interesting alternative solution. Alternative solutions to problems may make the whole Q\&A thread more useful to readers in the future. Increasing the perceived status of second or third answers to a question may get users post their first answer in a slower-pace environment.

Whether interventions are adapted from those designed for first-time questions or created particularly for the hurdle we study, our results suggest that different groups of people have significant differences in their tendency to post answers. This implies that some interventions may only work for certain kinds of users. For instance, if women are indeed less influenced by gamification, awarding more badges or tokens for posting a first answer is unlikely to help. This suggests why the leaky pipeline effect~\cite{shaw2018pipeline} may be so persistent.

Community culture also seems to matter. We know that in general users of online communities are much more likely to engage in more complex forms of participation if they feel like a member of a real social community and have social ties to other members that they repeatedly interact with~\cite{haythornthwaite2009online}. This may explain why users are less likely to post answers when they are active in larger communities. Previous work indicates that there are significant correlations in badges earned by users posting and commenting on the same posts~\cite{halavais2014badges}, suggesting that users are learning behavior from one another. It is also known that women are more likely to engage with a thread if they observe another woman~\cite{ford2017someone}. Though Stack Overflow does not have an explicit social network, users clearly respond differently when they recognize other people on threads. These hidden links may be valuable conduits for mentorship for potential new answer posters. In the future, these links could be made explicit through a system of referrals or collaborative answering. A core contributor could invite a less seasoned one to answer a question, promising to edit their first attempt. Rewarding such behavior could help many new users post their first answers.

\subsection{Limitations and Future Work}
We now turn to some limitations of our work and make suggestions how to address some of them in future research. Our analysis of gender differences has several limitations. Aside from the issue that gender is not a binary phenomenon, we recognize that our assumption that errors in the gender classification method are balanced is likely flawed. Research suggests that women are less likely to provide information about themselves in professional online platforms~\cite{altenburger2017there}. They may have good reason: research suggests that there are biases against identifiable women online, especially when they are underrepresented~\cite{bosu2019diversity, hidegender,sikdar2020effects}.

Likewise, our analysis of geography almost certainly suffers from non-random errors common to automated geocoding solutions~\cite{das2019gendered}. The tool we applied has a significant preference for English-language versions of places. The English-language orientation of the platform mitigates but does not remove this issue. Future work can certainly improve the geolocation of users, perhaps by including information collected from linked social media platforms. More broadly however, we must recognize that there are likely significant cultural and social biases in the population of users who reveal their location. In order to better integrate the impact of culture on user promotion~\cite{oliveira2016participation}, it is necessary to improve our inferences of user location.

One additional limitation of our analysis is that we ignore the role of social feedback plays in the integration of users into the community. For instance, a user posting a question which gets many up-votes and positive comments, may feel empowered to start answering questions sooner. We leave this analysis for future work, noting that it is difficult to measure the quality of a post, which is likely related with both receiving positive social feedback and the likelihood a user posts answers~\cite{burghardt2017myopia}.

Future work should certainly consider a more fluid notion of community than we do. People can and do post in many different tag communities on Stack Overflow. Experiences of seasoned users in new communities should provide useful insights into cultural and organizational differences between communities.

\subsection{Conclusion}
In this paper, we studied potential entry barriers to full participation in Stack Overflow focusing in particular on if and when active users begin posting answers. We found evidence that various individual features such as gender and tenure and community-level features like size or overall negativity are related to the likelihood and timing of a Stack Overflow user's first answer post. Our multiple regression models suggest that both level of features have an important relationship with user promotion along the pipeline of activity. Taken individually, the significant relationships between features and answer posting suggest that particular interventions may help certain groups of users. However, we also infer that the decision to start with posting answers has multi-faceted origins, and that there is not one obvious barrier blocking great participation across the board. In this way our work is only a first step in understanding why or why not users decide to move along the pipeline of contributions on online collaborative platforms.

\section*{Acknowledgements}
We thank Anna May, Kenny Joseph and Aleksandra Urman for comments and suggestions on a draft of this work. Aniko Hannak acknowledges funding from the Russell Sage Foundation (92-17-03).

\bibliography{sample-base}

\end{document}